\documentclass[a4paper, 10 pt, journal, twoside]{IEEEtran}  

\IEEEoverridecommandlockouts                              

 \usepackage[pdftex]{graphicx} 
\usepackage{amsmath,amssymb}
\usepackage{times}
\usepackage{bm}
\usepackage{here}
\usepackage{ascmac}
\usepackage{array}
\usepackage{longtable}
\usepackage{float}
\usepackage{wrapfig}
\usepackage{enumerate}
\usepackage{cite}
\usepackage{url}
\usepackage{multirow}
\usepackage{slashbox}
\usepackage{comment}
\usepackage{listings}
\usepackage{cases}
\usepackage{empheq}
\usepackage{nidanfloat}
\usepackage[subrefformat=parens]{subcaption}
\allowdisplaybreaks

\newcommand{\minimize}{\mathop{\rm minimize\hspace{3mm}}\limits}

\newtheorem{theorem}{{Theorem}}
\newtheorem{lemma}{{Lemma}}
\newtheorem{remark}{{Remark}}
\newtheorem{assumption}{{Assumption}}
\newtheorem{proof}{Proof}
\newcommand{\rfig}[1]{Fig.~\ref{#1}}

\newcommand{\R}{{\mathbb{R}}}
\newcommand{\Rn}{{\mathbb{R}}^n}
\newcommand{\Rm}{{\mathbb{R}}^m}
\newcommand{\rtheorem}[1]{Theorem~\ref{#1}}
\newcommand{\rlemma}[1]{Lemma~\ref{#1}}

\newcommand{\rappendix}[1]{Appendix~\ref{#1}}
\newcommand{\rassumption}[1]{Assumption~\ref{#1}}

\title{\LARGE \bf
Instant MPC for Linear Systems and Dissipativity-Based \\ Stability Analysis}

\author{Keisuke Yoshida, Masaki Inoue, and Takeshi Hatanaka
\thanks{K.~Yoshida and M.~Inoue contributed equally to this work.
}
\thanks{K.~Yoshida and M.~Inoue are with 
Department of Applied Physics and Physico-Informatics, Keio University; 
3-14-1 Hiyoshi, Kohoku-ku, Yokohama, Kanagawa 223-8521, JAPAN.
}       
\thanks{T.~Hatanaka is with 
Department of Systems and Control Engineering, Osaka University; 
2-1 Yamada-oka, Suita, Osaka 565-0871, JAPAN.
}
}

\begin{document}

\maketitle
\thispagestyle{empty}
\pagestyle{empty}

\begin{abstract}
This letter is devoted to the concept of {\it instant} model predictive control (iMPC) for linear systems.
An optimization problem is formulated to express the finite-time constrained optimal regulation control, like conventional MPC.
Then, iMPC determines the control action based on the optimization process rather than the optimizer, unlike MPC. 
The iMPC concept is realized by a continuous-time dynamic algorithm of solving the optimization;
 the primal-dual gradient algorithm is directly implemented as a dynamic controller.
On the basis of the dissipativity evaluation of the algorithm, the stability of the control system is analyzed.
Finally, a numerical experiment  is performed in order to demonstrate that iMPC emulates MPC 
 and to show its less computational burden.
\end{abstract}

\section{INTRODUCTION}
Model predictive control (MPC) is a control strategy in which 
an optimization problem is solved at each sampling instant and the control action is determined based on the optimizer.
Due to its high ability of handling various control requirements, 
MPC has attracted much attention.
See e.g., the survey paper \cite{mayne2014model} and references therein for the recent development of MPC.
There are still drawbacks of MPC;  due to its high computational burden,
implementing optimization especially in embedded controllers is not always viable.
To avoid destabilization, the terminal cost and constraints need to be chosen carefully \cite{mayne2000constrained}.
This letter addresses the computational burden and the stability guarantee by MPC.

There have been many trials of reducing the computational burden in the MPC framework 
\cite{wang2010fast,bemporad2002explicit,scokaert1999suboptimal,zeilinger2011real}.
The trials include the concepts of fast MPC \cite{wang2010fast}, explicit MPC \cite{bemporad2002explicit}, 
sub-optimal MPC \cite{scokaert1999suboptimal}, and their combination etc.
For example, in fast MPC or explicit MPC, the control law is designed offline, 
and the control action is determined online based on an implemented lookup table.
In sub-optimal MPC, 
the major effort of online computation is dedicated to finding a feasible solution for guaranteeing the stability, and 
the control action is determined by the initial feasible solution.
This letter proposes an alternative concept of MPC for reducing the computational burden;
the control action is determined {\it instantaneously} at the early stage of the optimization process rather than the optimizer. 
More specifically, 
the concept of such {\it instant} MPC (iMPC)
 is realized by the direct implementation of the primal-dual gradient algorithm \cite{arrow1958studies}, which 
is a {\it continuous-time} dynamic algorithm of pursuing the optimizer; 
the algorithm itself behaves as a dynamic controller.  Therefore, 
no optimization solver needs to be implemented, which would be benefitable for embedded controllers with limited computation performances and memory.

Recently, the primal-dual gradient algorithm has been studied well. 
For example, the stability, passivity, and control performance of the algorithm itself are addressed in \cite{feijer2010stability,cherukuri2016asymptotic,Qu_19,Yamamoto_12,Kosaraju_18,yamashita2018passivity,Simpson_18}.
Utilizing the passivity analysis \cite{Yamamoto_12,Kosaraju_18,yamashita2018passivity} in a more extended way,
we further evaluate the dissipativity \cite{hill1977stability,Brogliato_06} of the algorithm. 
Then, we derive the stability condition of the feedback control system composed of the algorithm and a plant system.

The rest of the letter is organized as follows.
Section I\hspace{-.1em}I formulates the problem statement of general MPC.
Section I\hspace{-.1em}I\hspace{-.1em}I gives the realization of iMPC and is dedicated to the dissipativity-based stability analysis of the control system.
Section I\hspace{-.1em}V gives a numerical experiment of iMPC.
Section V gives concluding remarks.
\smallskip

Notation: The symbol $\dot{v}$ represents the derivative  (upper Dini derivative) of a differentiable (continuous)  function $v(t)$. 

\section{Problem Statement}

\subsection{Plant System and Dissipativity}
We consider a continuous-time linear plant described by
\begin{align}
	\dot{\bm{x}}=A_c\bm{x}+B_c\bm{u}, \label{plant}
\end{align}
where $\bm{x}\in \Rn$ is the state, 
$\bm{u}\in\Rm$ is the input, and $A_c\in \R^{n\times n}$ and $B_c\in \R^{n\times m}$ are constant matrices.  
The following assumption is imposed on the plant system.
\smallskip

\begin{assumption}\label{QSR2}
The plant system (\ref{plant}) is {\it dissipative} with respect to the following $w_{(\ref{plant})}({\bm{x}},{\bm{u}})$
\begin{align*}
	w_{(\ref{plant})}({\bm{x}},{\bm{u}}) := \left[
	\begin{array}{c}
		{\bm{x}}\\
		{\bm{u}}
	\end{array}
	\right]^\top
	\left[
	\begin{array}{cc}
		Q_c & S_c \\
		S_c^{\top} & R_c
	\end{array}
	\right]
	\left[
	\begin{array}{c}
		{\bm{x}}\\
		{\bm{u}}
	\end{array}
	\right],
\end{align*}
where $Q_c\in \R^{n\times n}$, $R_c\in \R^{m\times m}$, and $S_c\in \R^{n\times m}$ are constant matrices; 
in other words, for some storage function ${{\cal S}}({\bm{x}})\geq 0$, it holds that
\begin{align*}
	\dot{{\cal S}}({\bm{x}})\leq w_{(\ref{plant})}({\bm{x}},{\bm{u}}).
\end{align*}
\end{assumption}
\medskip

The dissipativity property stated in the assumption is called {\it QSR dissipativity}.
The QSR dissipativity characterizes the set of various dynamical systems by the common matrix parameters $(Q_c, S_c, R_c)$.
In this letter, a control problem is addressed 
where control performance is pursued based on the detailed plant model (\ref{plant}), 
while the stability of the overall control system is guaranteed based on the set-characterizing parameters $(Q_c, S_c, R_c)$.

\subsection{Discrete-time MPC Framework}

In this subsection, 
the conventional framework of discrete-time MPC is reviewed.
To this end, we consider that the plant state $\bm{x}(t)$ is measured and the control input $\bm{u}(t)$ is applied to the plant
at time $t$. 
Then, the state behavior is predicted based on the discrete-time model
\begin{align*}
	\bm{x}_{(t,1)}&=A\bm{x}(t)+B\bm{u}(t),\\
	\bm{x}_{(t,p+1)}&=A\bm{x}_{(t,p)}+B\bm{u}_{(t,p)},\ \ \ p \in \{1,\, 2,\,...,\,N-1\},
\end{align*}
where system matrices $A$ and $B$ are given by
\[
	A:=e^{A_c \Delta t},\ \ \  B:=\int_0^{\Delta t} e^{A_c\tau} d\tau B_c,
\]
respectively.  
Here, 
$N$ represents the predictive horizon, and $\Delta t$ is the time step for the discretization.
Then, we formulate an optimization problem associated with constrained optimal control including the model-based state prediction:
\begin{subequations}\label{optpro}
\begin{align}
	&\displaystyle \minimize_{\bm{z}\in\R^{(m+n)N}}{f(\bm{z})}\label{optpro1}\\
	&\mathrm{subject\ to}\hspace{3mm}g(\bm{z}) \leq0,\label{optpro2}\\
	&\hspace{18.5mm}h(\bm{z},\bm{x}(t)):=H\bm{z}+V\bm{x}(t)=0,\label{optpro3}
\end{align}
\end{subequations}
where $\bm{z}\in\R^{(m+n)N}$ is the decision variable defined by
\[
	\bm{z}=[\, \bm{u}{(t)}^\top\, \bm{u}_{(t,1)}^\top\, \cdots \, \bm{u}_{(t,N-1)}^\top\ \bm{x}_{(t,1)}^\top\, \cdots \, \bm{x}_{(t,N)}^\top\,]^{\top}.
\]

The symbol $f$ is the cost function, which includes the stage cost and the terminal cost \cite{mayne2000constrained}.
We emphasize here that in MPC setup the equality constraint (\ref{optpro3}) includes the sequence of the discretized plant models; 
in other words, the matrices $H$ and $V$ of (\ref{optpro3}) are expressed as follows;
{\arraycolsep = 2.5pt
\begin{align*}
&H=
\left[\begin{array}{c}
\begin{array}{cccccccccc}
B&0&&\cdots&0&-I&0&&\cdots&0 \\ 
0&B&0&\cdots&0&A&-I&0&\cdots&0\\
 & & & &\vdots& & & & & \\
0&&\cdots&0&B&0&\cdots&0&A&-I
\end{array}
\\\hline
\ast
\end{array}
\right],\\
&V=\left[
\begin{array}{cccc|c}
A^\top & 0 & \cdots &0 &\ast
\end{array}
\right]^\top,
\end{align*}
}%
where $\ast$ represents matrices come from some requirements by control designer.

Conventional MPC schemes are based on the implicit assumption that 
the problem (\ref{optpro}) is solved in some time interval such as the sampling period. 
Then, MPC outputs $\bm{u}^{\ast}(t)$, which is the first $m$ elements of the optimizer $\bm{z}^{\ast}$, 
to the plant system as a control action.
Although  there are a variety of numerically tractable methods of solving convex optimization \cite{Rao_98,Kirches_11,Richter_12,Patrinos_14,Pu_17}, 
they still require a high computational load annoying control designers.

This letter addresses the drawback of conventional MPC 
and aims at realizing the idea of ``instantaneously'' deciding the control action $\bm{u}(t)$ 
even in the MPC framework.

The following assumptions are imposed on the optimization problem (\ref{optpro}).
\smallskip
\begin{assumption}\label{fg}
The function $f$ is strongly convex and continuously differentiable.  
The function $g$ is convex and continuously differentiable. 
For any $\bm{x}$, there exists $\bm{z}$ such that $g(\bm{z}) < 0$, $h(\bm{z},\bm{x}) = 0$, and $f(\bm{z})$ is finite.
In addition, the gradients $\nabla f$ and $\nabla g$ are locally Lipschitz and satisfy $\nabla f({0})={0}$.
\end{assumption}


\section{Instant Model Predictive Control}

\begin{figure}[t]
 \begin{center}
 \includegraphics[width=0.8\hsize]{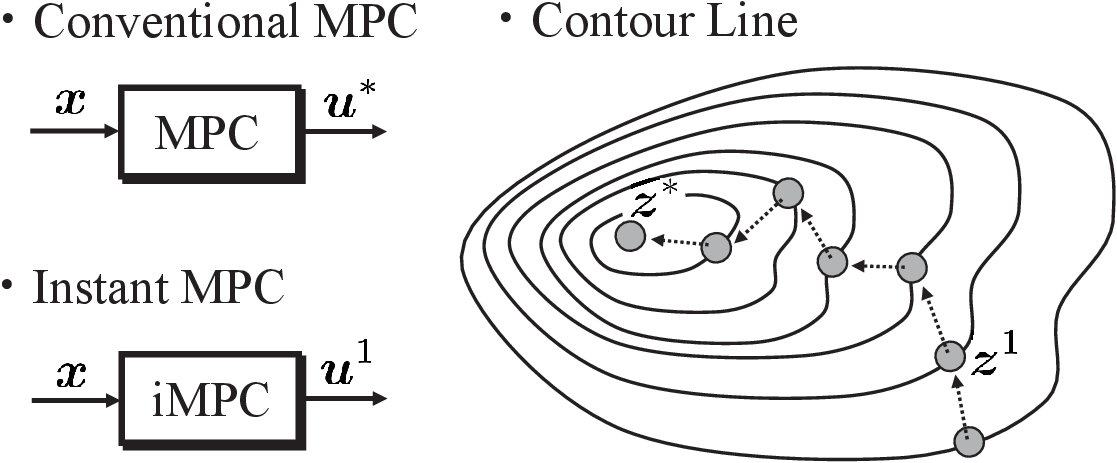}
 \caption{Sketch of control action decisions by MPC and iMPC.
The figure in the right-side illustrates optimization process on the contour lines of a cost function. 
}%
 \label{image}
 \end{center}
 \end{figure}

{First, we introduce the brief idea of  {\it instant} MPC (iMPC) in a discrete-time algorithm of solving (\ref{optpro}).
Let the optimizing sequence be generated by the algorithm and denoted by $\{\bm{z}^1,\bm{z}^2,\ldots \}$.
In conventional MPC, the control input $\bm{u}(t)$ is determined based on the optimizer, 
which is denoted by $\bm{z}^\ast =\bm{z}^i$, $i\to \infty$ and  is obtained by e.g., the gradient method.
On the other hand, in iMPC, $\bm{u}(t)$ is {\it instantaneously}  determined based on $\bm{z}^1$,
which is the first element of the optimizing sequence. 
The  iMPC concept is illustrated in \rfig{image}.
It is expected that iMPC contributes to significantly reducing the computational burden.
}

The rest of this section is organized as follows.
First, we present a realization of the iMPC concept, which is stated by a {\it discrete-time} algorithm above,  
by a {\it continuous-time} algorithm of solving (\ref{optpro}), in particular, the primal-dual gradient algorithm \cite{arrow1958studies,feijer2010stability,cherukuri2016asymptotic}.  
Then, the overall control system is expressed by a state equation.
Next, we analyze an equilibrium point of the control system to guarantee its stability based on dissipativity theory \cite{hill1977stability,Brogliato_06}.
Finally, we give some remarks on iMPC including ideas of further extensions and of practical implementation.

\subsection{Realization of Instant MPC}

Under \rassumption{fg}, for any given $\bm{x}$, the optimization problem (\ref{optpro}) has a unique optimizer 
$(\bm{z}^{\ast},\bm{\mu}^{\ast},\bm{\lambda}^{\ast}) = (\bm{z}^{\ast}(\bm{x}),\bm{\mu}^{\ast}(\bm{x}),\bm{\lambda}^{\ast}(\bm{x}))$.  
It is well known that the optimizer satisfies the following KKT condition \cite{boyd2004convex}:
\begin{align*}
	&\nabla f(\bm{z}^{\ast})+\nabla g(\bm{z}^{\ast})\bm{\mu}^{\ast} +\nabla h(\bm{z}^\ast,\bm{x})\bm{\lambda}^{\ast}=0,\\
	&\bm{\mu}^{\ast}\geq 0,\hspace{3mm}g(\bm{z}^{\ast})\leq 0,\hspace{3mm}\bm{\mu}^{\ast}\circ g(\bm{z}^{\ast})=0,\\
	&h(\bm{z}^{\ast},\bm{x})=0,
\end{align*}
where the symbol $\circ$ represents the Hadamard product.
In iMPC, we are particularly interested in solution methods of 
pursuing the optimizer $(\bm{z}^{\ast}(\bm{x}(t)),\bm{\mu}^{\ast}(\bm{x}(t)),\bm{\lambda}^{\ast}(\bm{x}(t)))$ for time-varying $\bm{x}(t)$.
In this letter, we consider the following primal-dual gradient algorithm as a solution method.
\begin{subequations}\label{pdprop}
\begin{align}
	&\dot{\bm{z}}=-\nabla f(\bm{z})-\nabla g(\bm{z})\bm{\mu} -K\nabla h(\bm{z},\bm{x})(\bm{\lambda}+\beta\dot{\bm{\lambda}}),\label{pdprop1}\\
	&\dot{\bm{\mu}}=[g(\bm{z})]^{+}_{\bm{\mu}},\label{pdprop2}\\
	&\dot{\bm{\lambda}}=-{\alpha}({1+\alpha\beta})^{-1}\bm{\lambda} + ({1+\alpha\beta})^{-1} h(\bm{z},\bm{x}),\label{pdprop3}
\end{align}
\end{subequations}
where $\alpha$ is a positive constant, $\beta$ is a nonnegative constant, $K$ is given by 
\begin{align}
	K =1+2\alpha\beta, \label{K}
\end{align}
and $[\cdot]^{+}_{\ast}$ is an operator defined as
\begin{align*}
	[\sigma]^{+}_{\varepsilon}:=
	\begin{cases}
	\sigma,\hspace{18mm}\mathrm{if}\ \varepsilon>0 \\
	\text{max}\{0,\sigma\},\hspace{5mm}\mathrm{if}\ \varepsilon=0.
	\end{cases}
\end{align*}
for scalars $\sigma,\varepsilon\in\R$.
For vectors $\sigma,\varepsilon\in\R^n$, 
$[\sigma]^{+}_{\varepsilon}$ denotes the vector whose $i$-th component is $[\sigma_i]^{+}_{\varepsilon_i},i\in\{1,...,n\}$.

The algorithm (\ref{pdprop}) is modified from the original one proposed in \cite{cherukuri2016asymptotic};
the parameters $\alpha$ and $\beta$ are additionally introduced to the original algorithm in order to reinforce  its ``dissipativity''.  
The reinforcement plays key roles in guaranteeing the stability of the control system and 
also in reducing the gap between iMPC and MPC. 
The details of the roles are mentioned later in Remark \ref{rem:alpha} and utilized in Section \ref{sec:sim}.

In this letter, the iMPC concept is realized by  the {\it continuous-time} algorithm (\ref{pdprop}). 
In the realization, we let
\begin{align}
	\bm{u}(t) = E \bm{z}(t), \label{iMPCimple}
\end{align}
where 
\[
	E:= [\, I\ 0\,\cdots\, 0\,]\in\R^{m\times (m+n)N}.
\] 
Then, the overall control system $\Sigma_{\rm all}$ is composed of the plant (\ref{plant}) and the {\it controller} (\ref{pdprop}) and (\ref{iMPCimple}).
In the control system,  the plant state $\bm{x}(t)$ is measured from  (\ref{plant}) and fed back to (\ref{pdprop}) continuously, 
and the control input $\bm{u}(t)$ is generated by (\ref{pdprop}) and actuates (\ref{plant}) continuously.

\subsection{Equilibrium Analysis}

\begin{figure*}[t]
 \begin{center}
 \includegraphics[scale=0.4]{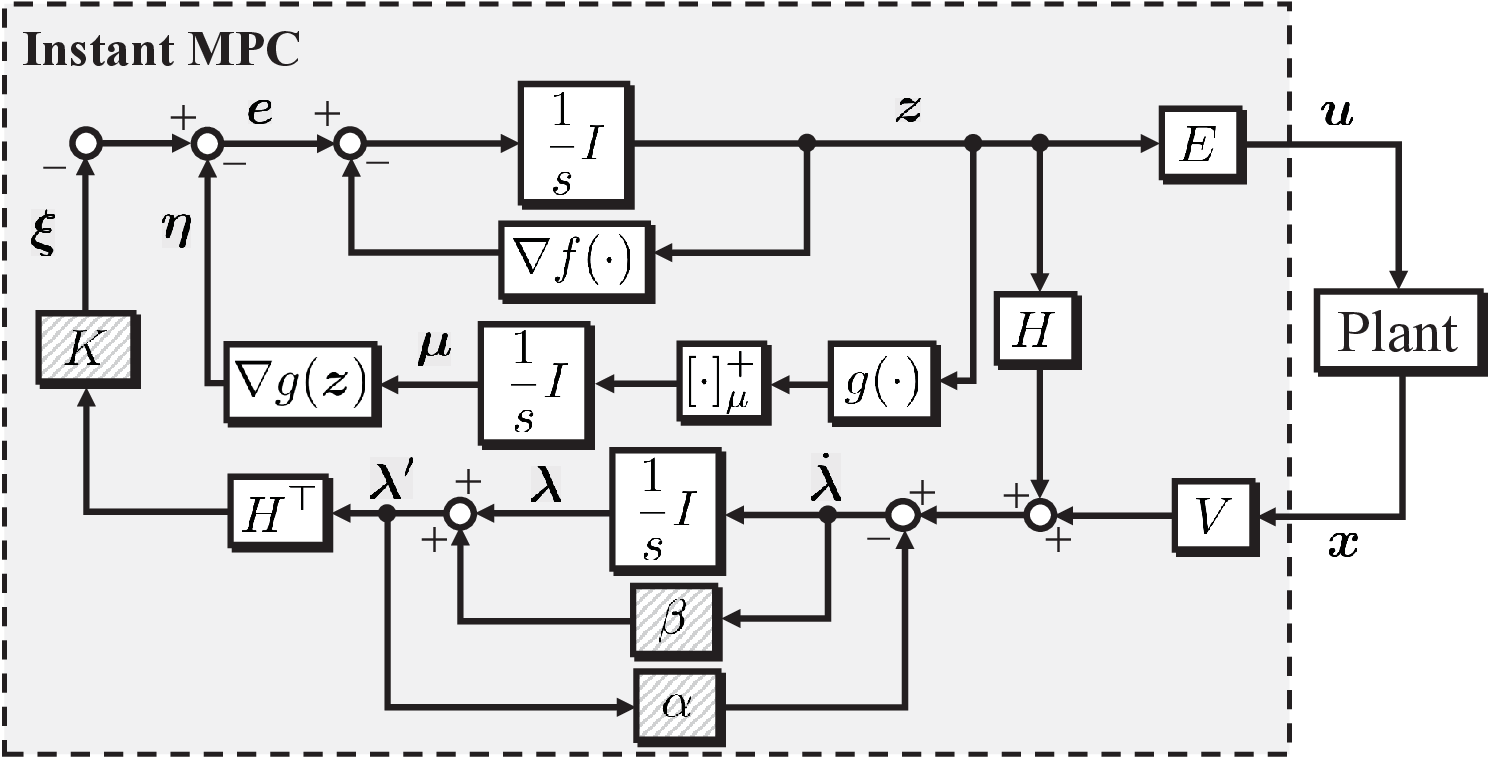}
 \caption{Block diagram of overall control system $\Sigma_{\rm all}$.}%
 \label{WIMPC}
 \end{center}
 \end{figure*}

In this subsection, we analyze the stability of the overall control system $\Sigma_{\rm all}$.
The block diagram of $\Sigma_{\rm all}$ is illustrated in \rfig{WIMPC}, 
where $\bm{\xi}$, $\bm{\eta}$, $\bm{e}$, and $\bm{\lambda}^\prime$ are defined by
\begin{align}
	\bm{\xi}&:=K \nabla h(\bm{z},\bm{x})\bm{\lambda}^\prime = K H^\top \bm{\lambda}^\prime,\label{xi}\\
	\bm{\eta}&:=\nabla g(\bm{z})\bm{\mu},\label{eta}\\
	\bm{e}&:=-\bm{\xi}-\bm{\eta},\label{e}\\
	\bm{\lambda}^\prime&:=\bm{\lambda}+\beta\dot{\bm{\lambda}},\label{lambda}
\end{align}
respectively.

We aim at finding the equilibrium of $\Sigma_{\rm all}$.  
The equilibrium $(\bm{x}^{e},\bm{z}^{e},\bm{\mu}^{e},\bm{\lambda}^{e})$ satisfies 
\begin{subequations}\label{KKT2}
\begin{align}
	&A_c\bm{x}^{e}+B_c\bm{u}^{e}={0},\\
	&\nabla f(\bm{z}^{e})+\nabla g(\bm{z}^{e})\bm{\mu}^{e} +KH^{\top}\bm{\lambda}^{e}={0},\\
	&\bm{\mu}^{e}\geq {0},\hspace{3mm}g(\bm{z}^{e})\leq {0},\hspace{3mm}\bm{\mu}^{e}\circ g(\bm{z}^{e})={0},\\
	&H\bm{z}^{e}+V\bm{x}^{e}-\alpha \bm{\lambda}^{e}={0}.
\end{align}
\end{subequations}
Recalling $\nabla f({0})={0}$ from \rassumption{fg}, 
we see that 
$(\bm{x}^{e},\bm{z}^{e},\bm{\mu}^{e},\bm{\lambda}^{e})=(0,0,0,0)$ satisfies (\ref{KKT2}).
This fact is stated in the following lemma.  
\smallskip

\begin{lemma}\label{OSP}
The origin is an equilibrium of $\Sigma_{\rm all}$.
\end{lemma}
\smallskip

Note that the uniqueness of the equilibrium is not stated in the lemma.
The stability of the origin is analyzed in the next subsection.

\subsection{Dissipativity of Instant MPC and Stability Assurance}

In this subsection, the stability of the overall control system is analyzed 
based on dissipativity theory \cite{hill1977stability,Brogliato_06}.  
Recall that  the dissipativity property of the plant system is assumed in (\ref{QSR2}) and characterized by $(Q_c, S_c, R_c)$.
We then aim at evaluating the dissipativity of the iMPC realization (\ref{pdprop}) as follows.
\smallskip

\begin{lemma}\label{QSR1}
Let $\rho$ be a positive constant such that
\begin{align}
	\nabla f(\bm{z})^{\top}\bm{z}
	\geq
	\rho \bm{z}^\top \bm{z} 
	\label{strongconvex}
\end{align}
holds for all $\bm{z} \in \R^{(m+n)N}$.
Then, the iMPC (\ref{pdprop}) is dissipative with respect to the following supply rate:
{\arraycolsep = 2.5pt
\begin{align}\nonumber
&w_{(\ref{pdprop})}(\bm{z},\bm{x})\\ 
&:=\left[\begin{array}{c}
	{\bm{z}}\\
	{\bm{x}}
\end{array}\right]^\top
\left[
\begin{array}{cc}
-\rho I -\beta H^\top H & -\beta H^\top V \\
-\beta V^\top H & \frac{1}{4\alpha(1+\alpha\beta)}A^{\top}A 
\end{array}
\right]
\left[\begin{array}{c}
	{\bm{z}}\\
	{\bm{x}}
\end{array}\right].
\label{qsr11}
\end{align}
}
\end{lemma}
\smallskip

\begin{proof}
See \rappendix{prQSR1}.
\end{proof}
\smallskip

There have been various studies on the property of the primal-dual gradient algorithm.
For example, the asymptotic stability, passivity, and control performance of the isolated algorithm with no connection to plant systems, are studied in e.g.,
\cite{feijer2010stability,cherukuri2016asymptotic,Qu_19,Yamamoto_12,Kosaraju_18,yamashita2018passivity,Simpson_18}.
It is noted that the input-output ports considered in Lemma \ref{QSR1} are different from those in the previous works.
The dissipativity is evaluated for the input-output pair ($\bm{x}$, $\bm{z}$) in Lemma \ref{QSR1}, 
while the passivity is shown for ($\bm{x}$, $V^{\top}\bm{\lambda}^\prime$) or other ``symmetric'' pairs in e.g., \cite{yamashita2018passivity}.

Here, we recall that the control input is given by (\ref{iMPCimple}) and that
the plant system (\ref{plant}) generates the state $\bm{x}(t)$ continuously. 
From \rassumption{QSR2}, the dissipativity of the plant (\ref{plant}) with (\ref{iMPCimple}) is evaluated as follows.
Letting the storage function be given by 
	${\cal S}_{(\ref{plant},\ref{iMPCimple})}({\bm{x}}):=\frac{1}{2}{\bm{x}}^\top {\bm{x}}$,
we show that the input-output system from $\bm{z}$ to $\bm{x}$ is dissipative with respect to the following supply rate:
\begin{align}
w_{(\ref{plant},\ref{iMPCimple})}({\bm{x}},{\bm{z}}) := 
\left[\begin{array}{c}
	{\bm{x}}\\
	{\bm{z}}
\end{array}\right]^\top
\left[\begin{array}{cc}
	Q_c & S_cE \\
	E^{\top}S_c^{\top} & E^{\top}R_cE
\end{array}\right]
\left[\begin{array}{c}
	{\bm{x}}\\
	{\bm{z}}
\end{array}\right].\label{qsr12}
\end{align}

On the basis of the dissipativity evaluation given by (\ref{qsr11}) and (\ref{qsr12}),
we derive a condition for the {\it partial stability} of the overall control system $\Sigma_{\rm all}$.
\smallskip

\begin{theorem}\label{main}
Suppose that 
\begin{align*}
	&{Q}_{\rm all}:=\\
	&\left[
	\begin{array}{cc}
	-\rho I -\beta H^\top H +\delta E^{\top}R_c E & -\beta H^\top V + \delta E^{\top}S_c^{\top}\\
	-\beta V^\top H + \delta S_cE & \frac{1}{4\alpha(1+\alpha\beta)}A^{\top}A +\delta Q_c
	\end{array}
	\right]
	\label{mainQ}
\end{align*}
is negative definite for some positive constant $\delta$.
Then, the origin $(\bm{x}, \bm{z},\bm{\mu},\bm{\lambda})=(0,0,0,0)$ 
is asymptotically stable with respect to $(\bm{x},\bm{z},\bm{\lambda})$; i.e., 
the origin is 
Lyapunov stable, 
and for any initial state $(\bm{x}(0), \bm{z}(0),\bm{\mu}(0),\bm{\lambda}(0))$, 
it holds that $(\bm{x}(t),\bm{z}(t),\bm{\lambda}(t)) \rightarrow (0,0,0)$, $t \rightarrow \infty$.
\end{theorem}
\smallskip

\begin{proof}
See \rappendix{prmain}.
\end{proof}
\smallskip

{
The result of this letter is compared with the related literature.
In  \cite{Jokic_09,Lawrence_18,Nicotra_18}, 
continuous-time algorithms of solving optimization problems are embedded into dynamic controllers,  
called dynamic KKT control \cite{Jokic_09}, optimal steady-state control \cite{Lawrence_18}, and 
dynamically embedded MPC \cite{Nicotra_18}.  
The main difference from them, i.e., the contribution of this letter, is 
dissipativity-based stability analysis as presented in \rtheorem{main}. 
We note that  the stability-guaranteeing parameters $(\alpha, \beta)$ are characterized in the theorem.
Then, based on the set of the parameters,  
control designer can pursue the control performance without impairing the stability. 
The dissipativity-based stability analysis in iMPC releases 
the troublesome choice of the terminal cost or terminal state constraints, which needs to be addressed essentially in conventional MPC  \cite{mayne2000constrained}.
In iMPC, there is no explicit requirement on the cost or constraint for guaranteeing the stability of $\Sigma_{\rm all}$ unlike conventional MPC.
}

\smallskip

\begin{remark}
\label{rem:alpha}{(Role of $\alpha$)}\ 
The role of $\alpha$, introduced  in the proposed algorithm (\ref{pdprop}), is discussed in this remark. 
Let $\alpha$ be sufficiently small to reduce the algorithm to the conventional one \cite{cherukuri2016asymptotic} with fixed $\beta$.
Then, $Q_{\rm all}$ of \rtheorem{main} cannot be negative definite, and the stability is not guaranteed by the dissipativity-based analysis.
This concludes that $\alpha$ plays a role in guaranteeing the stability of $\Sigma_{\rm all}$. 
\end{remark}
\smallskip

\begin{remark}
{(Robustness)}\ 
It should be noted that the continuous-time model (\ref{plant}) is not utilized directly for guaranteeing the stability by \rtheorem{main}.
The stability $\Sigma_{\rm all}$ is not impaired as long as $Q_{\rm all}$ is negative definite, which is independent of $(A_c,B_c)$.
To be extreme, \rtheorem{main} is still valid even if $(A,B)$ and $(A_c,B_c)$ are completely unrelated or 
if the actual plant system is nonlinear instead of (\ref{plant}).
This exhibits the robustness of the proposed iMPC to modeling errors. 
In iMPC, the plant model contributes directly  to improving control performance.
\end{remark}

\subsection{Extensions of Algorithm}

In this subsection, two extensions of the algorithm (\ref{pdprop}) are discussed.
\smallskip

\subsubsection{Algorithm with Projection}
We recall that iMPC realized by (\ref{pdprop}) and (\ref{iMPCimple}) outputs the transient response of $\bm{u}(t)$.
Then, the inequality and equality constraints, given by (\ref{optpro2}) and (\ref{optpro3}),  may not be satisfied except the steady state.
To activate the constraints for all time, we can apply the projection of $\bm{z}$ onto the constraints.
For example, the projection onto the equality constraint (\ref{optpro3}) is described by 
\begin{align*}
	\bm{z}_{\rm proj}=\{ I-H^{\top}(HH^{\top})^{-1}H \} \bm{z}-H^{\top}(HH^{\top})^{-1}V\bm{x}.
\end{align*}
The projection-based iMPC is realized by $\bm{u} = E \bm{z}_{\rm proj}$ instead of (\ref{iMPCimple}).
The projection onto the inequality constraint (\ref{optpro2}) requires of solving some optimization problem.
The details of handling inequality constraints are omitted in this letter, due to the severe limitation of pages.
\smallskip

\subsubsection{Algorithm with Additional Parameters}

The algorithm (\ref{pdprop}) contains two design parameters  $\alpha$ and $\beta$, 
which can contribute to stability assurance and improving control performance.
An extension of the algorithm with additional design parameters is given as follows.
Letting $\gamma$ be a non-negative constant, we define the algorithm 
\begin{subequations}\label{gamma}
\begin{align}
	&\dot{\bm{z}}=-\nabla f(\bm{z})-\nabla g(\bm{z}+\gamma \dot{\bm{z}})\bm{\mu} -H^{\top}(\bm{\lambda}+\beta\dot{\bm{\lambda}}),\\
	&\dot{\bm{\mu}}=[g(\bm{z}+\gamma \dot{\bm{z}})]^{+}_{\bm{\mu}},\\
	&\dot{\bm{\lambda}}=H(\bm{z}+\gamma \dot{\bm{z}})+V\bm{x}-\alpha (\bm{\lambda}+\beta\dot{\bm{\lambda}}).
\end{align}
\end{subequations}
We see that the algorithm (\ref{gamma}) is reduced to (\ref{pdprop}) if $\dot{\bm{z}}= 0$ holds.  
This fact motivates us to replace (\ref{pdprop}) by this (\ref{gamma}) 
aiming at improving the transient response of $\Sigma_{\rm all}$, while preserving its steady state property. 


\section{Numerical Experiment}\label{sec:sim}

In this section, the effectiveness of the proposed iMPC scheme is presented, 
and the role of the design parameter $\beta$ is demonstrated.  
To focus only on the behavior of iMPC and its computational burden, 
inequality constraints are not considered in the numerical experiment.


Consider the DC motor model given in \cite{matlabLTI}, where 
the state $\bm{x}=[\,i\ \, \omega\,]^{\top}$ is composed of the current $i$ and the angular velocity $\omega$.
The control input $\bm{u}$ is the applied voltage.
The system dynamics are described by the continuous-time model (\ref{plant}) with system matrices
\begin{align*}
A_c=
\left[
\begin{array}{cc}
-4 & -0.03 \\
0.75 & -10 
\end{array}
\right]
\hspace{3mm}
B_c=
\left[
\begin{array}{c}
2 \\
0 
\end{array}
\right].
\end{align*}
The control objective is to track the desired reference $r=[\,i_r\ \omega_r\,]^{\top}$, where $i_r=200/3$ and $\omega_r=5$.
To this end, we formulate an optimization problem where the predictive horizon is $N=30$, the sampling time is $\Delta t=0.1s$, and the cost function is given by
$f(\bm{z})=\bm{z}^\top F \bm{z}$, where $F={\rm blkdiag}(I_{N},1000I_{2N})$.
The state responses and computational burden are examined 
by applying 1) conventional MPC, 2) iMPC with $(\alpha,\beta)=(10,10)$, and 3) iMPC with $(\alpha,\beta)=(10,1000)$.  
{
The computational burden is evaluated by the processing time for each control input decisions.  
The algorithms of MPC and iMPC are implemented in CPU {[Core i7-6500U 2.5 GHz].}}
For Case 1, the quadratic optimization problem is solved using the Matlab {\tt quadprog} intentionally, though its explicit solution is available. 
For Cases 2 and 3, the algorithm (\ref{pdprop}) is implemented to pursue the optimal solution.  
In Case 2, the stability of the control system is guaranteed in advance by \rtheorem{main}; 
we show that $Q_{\rm all}$ is negative definite by letting $Q_c=A_c$, $S_c=\frac{1}{2}B_c$, $R_c=0$, $\rho=2$, and $\delta=1$.
On the other hand, in Case 3, $Q_{\rm all}$ cannot be negative definite.  
The stability is not guaranteed in advance, but is shown by the numerical experiment below.
\smallskip

\begin{figure}[t]
 \begin{center}
 \includegraphics[scale=0.6]{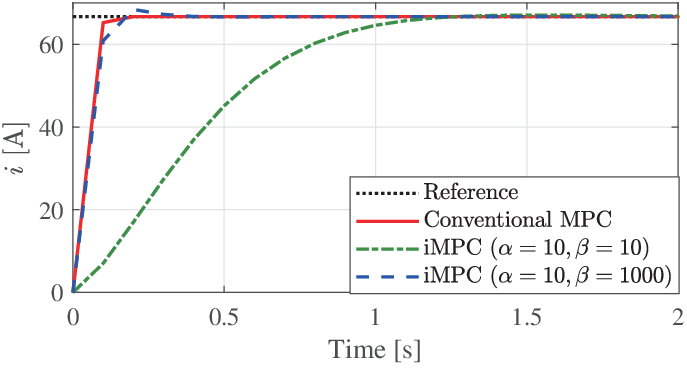}
 \includegraphics[scale=0.6]{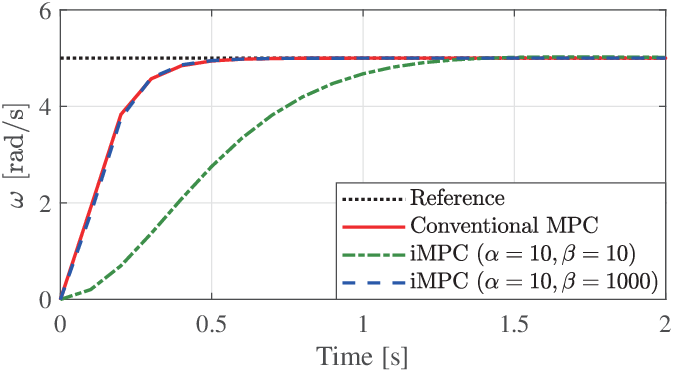}
 \caption{State response $[\, i\ \omega\,]$.}%
 \label{Response}
 \end{center}
 \begin{center}
 \includegraphics[scale=0.6]{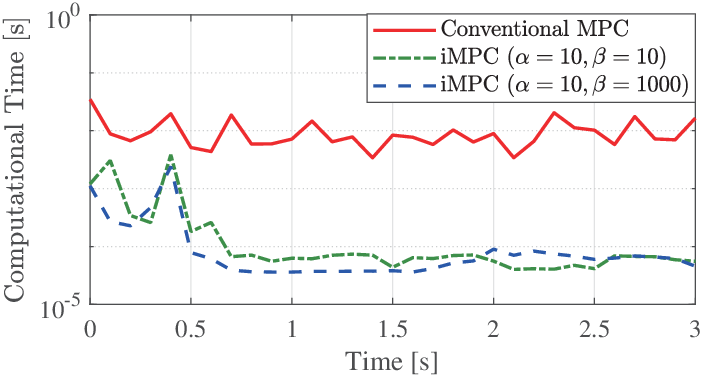}
 \caption{Computational time (in logscale). }%
 \label{Burden}
 \end{center}
 \end{figure}

The results of the state response and processing time are given in Figs.~\ref{Response} and \ref{Burden}, respectively. 
In the figures, the results of Cases 1, 2, and 3 are depicted by red solid, green dot-dashed, and blue dashed lines, respectively.
We see from Fig.~\ref{Response} that every state trajectory converges to the reference value.
Although the tracking performance is deteriorated in iMPC compared with MPC, 
tuning of $(\alpha, \beta)$ significantly relax the deterioration.  
In fact, the response of iMPC in Case 3 is almost the same as that of MPC in Case 1.
The average processing time is $9.80 {\rm ms}$ for Case 1, $0.331 {\rm ms}$ for Case 2, and $0.185 {\rm ms}$ for Case 3.
We conclude that iMPC contributes to reducing the computational burden.

\section{Concluding Remarks}

This letter was devoted to the concept of {\it instant} MPC (iMPC), which dynamically emulated conventional MPC.
To realize the concept, the continuous-time {\it dynamic} algorithm of solving optimization problems, 
in particular, the primal-dual gradient algorithm,
replaced the conventional {\it static} predictive controller. 
The iMPC realization contributed to significantly reduce the computational burden. 
The stability of the overall control system was analyzed based on dissipativity theory, 
where any requirement on the terminal cost or constraints were not necessary unlike conventional MPC.

The concept of iMPC  includes a variety of extensions and applications.
Alternative algorithm replacing (\ref{pdprop}) will be studied  and applied for tracking control.
In the problem setting of this letter, it is assumed that the plant system is stable or stabilized in advance by some local controller.
Stabilizing iMPC will be addressed for unstable plant systems. 
As implied in Case 3 of Section \ref{sec:sim}, 
\rtheorem{main} provides only a sufficient condition for the stability.
Less conservative stability analysis will be addressed in future work.  

\bibliographystyle{IEEEtran}
\bibliography{ref_MI}

\appendix

\subsection{Proof of \rlemma{QSR1}}\label{prQSR1}
{
Let the storage function of (\ref{pdprop1}) be given by 
${\cal S}_{(\ref{pdprop1})}({\bm{z}}):=\frac{1}{2}{\bm{z}}^\top {\bm{z}}$.
In addition, noting (\ref{xi}), (\ref{eta}), and (\ref{e}),  
we rewrite (\ref{pdprop1}) by $\dot{\bm{z}} = -\nabla f(\bm{z}) + e$.
Then, the time derivative of ${\cal S}_{(\ref{pdprop1})}({\bm{z}})$ along the state trajectories is given by
\begin{align*}
\dot{\cal  S}_{(\ref{pdprop1})}({\bm{z}})
&=\bm{e}^{\top}\bm{z}-\nabla f(\bm{z})^{\top}\bm{z}.
\end{align*}
From \rassumption{fg}, there exists a positive constant $\rho$ such that (\ref{strongconvex}) holds
for all $\bm{z} \in \R^{(m+n)N}$.
Then, it follows that
\begin{align*}
\dot{\cal S}_{(\ref{pdprop1})}({\bm{z}})
\leq 
\bm{e}^{\top}\bm{z}- \rho \bm{z}^{\top}\bm{z}.
\end{align*}

It is known that the system (\ref{pdprop2}) is passive from $\bm{z}$ to $\bm{\eta}$ with respect to the storage function
 ${\cal S}_{(\ref{pdprop2})}(\bm{\mu}):=\frac{1}{2}{\bm{\mu}}^\top {\bm{\mu}}$ \cite{hatanaka2018passivity}, i.e., the following inequality holds.
\begin{align*}
\dot{\cal S}_{(\ref{pdprop2})}(\bm{\mu})
\leq
\bm{\eta}^{\top}\bm{z}.
\end{align*}   

Furthermore, let the storage function of (\ref{pdprop3}) be given by 
${\cal S}_{(\ref{pdprop3})}({\bm{\lambda}}):=\frac{1}{2}{\bm{\lambda}}^\top {\bm{\lambda}}$.
Then, its time derivative along the state trajectories is given by
\begin{align*}
	\dot{\cal S}_{(\ref{pdprop3})}({\bm{\lambda}})
	=&-\frac{\alpha}{1+\alpha\beta}\bm{\lambda}^{\top} \bm{\lambda} + \frac{1}{1+\alpha\beta}\bm{\lambda}^\top h(\bm{z},\bm{x}).
\end{align*}
Here, recalling from (\ref{pdprop3}) and (\ref{lambda}), we have 
\begin{align*}
	\bm{\lambda} &= \bm{\lambda}^\prime - \beta \dot{\bm{\lambda}}\\
								& = (1+\alpha\beta) \bm{\lambda}^\prime - \beta h(\bm{z},\bm{x}).
\end{align*}
It follows from (\ref{optpro3}) and (\ref{K}) that 
\begin{align*}
	\dot{\cal S}_{(\ref{pdprop3})}({\bm{\lambda}})
	=&-\alpha(1+\alpha\beta){\bm{\lambda}^\prime}^{\top}{\bm{\lambda}}^\prime+K{\bm{\lambda}^\prime}^{\top}H\bm{z}\\
	&+K{\bm{\lambda}^\prime}^{\top}V\bm{x}-\beta h(\bm{z},\bm{x})^\top h(\bm{z},\bm{x}).
\end{align*}
Letting
 $\tilde{\bm{\lambda}}:=\bm{\lambda}^\prime-\frac{(1+2\alpha\beta)}{2\alpha(1+\alpha\beta)}V\bm{x}$ and 
noting (\ref{xi}),
we obtain
\begin{align*}
\dot{\cal S}_{(\ref{pdprop3})}({\bm{\lambda}})
=&-\alpha(1+\alpha\beta)\tilde{\bm{\lambda}}^{\top}\tilde{\bm{\lambda}}
+\frac{(1+2\alpha\beta)^2}{4\alpha(1+\alpha\beta)}\bm{x}^{\top}V^{\top}V\bm{x}\\
&+K {\bm{\lambda}^\prime}^{\top}H\bm{z}-\beta h(\bm{z},\bm{x})^\top h(\bm{z},\bm{x})\\
\leq 
&\frac{(1+2\alpha\beta)^2}{4\alpha(1+\alpha\beta)}\bm{x}^{\top}A^{\top}A\bm{x}+\bm{\xi}^{\top}\bm{z}\\
&-\beta h(\bm{z},\bm{x})^\top h(\bm{z},\bm{x}).
\end{align*}

Finally, let the storage function of (\ref{pdprop}) be given by 
\[
	{\cal S}_{(\ref{pdprop})}(\bm{z},\bm{\mu},\bm{\lambda}):={\cal S}_{(\ref{pdprop1})}({\bm{z}})+{\cal S}_{(\ref{pdprop2})}(\bm{\mu})+{\cal S}_{(\ref{pdprop3})}({\bm{\lambda}}).
\]
Then, we have 
\begin{align*}
\dot{\cal S}_{(\ref{pdprop})}(\bm{z},\bm{\mu},\bm{\lambda})
\leq
&-\rho \bm{z}^{\top}\bm{z}
+\frac{(1+2\alpha\beta)^2}{4\alpha(1+\alpha\beta)}\bm{x}^{\top}A^{\top}A\bm{x}\\
&-\beta h(\bm{z},\bm{x})^\top h(\bm{z},\bm{x}).
\end{align*}
From (\ref{optpro3}), we see that 
the statement of the lemma holds.
\smallskip

\subsection{Proof of \rtheorem{main}}\label{prmain}

The candidate of a Lyapunov function is given by 
\begin{align*}
	{\cal V}(\bm{x}_k,\bm{z},\bm{\mu},\bm{\lambda})={\cal S}_{(\ref{pdprop})}(\bm{z},\bm{\mu},\bm{\lambda}) + \delta {\cal S}_{(\ref{plant},\ref{iMPCimple})}({\bm{x}}),
\end{align*}
which is a positive definite function. 
From (\ref{qsr11}) and (\ref{qsr12}), it follows that
\begin{align}
	\dot{\cal V}(\bm{x},\bm{z},\bm{\mu},\bm{\lambda})\leq
	\left[
	\begin{array}{c}
	\bm{z}\\
	\bm{x}
	\end{array}
	\right]^\top
		Q_{\rm all}
	\left[
	\begin{array}{c}
	\bm{z}\\
	\bm{x}
	\end{array}
	\right]=: q(\bm{x},\bm{z}).
\label{prooflast}
\end{align}
Then, 
the origin $(\bm{x},\bm{z},\bm{\mu},\bm{\lambda})=(0,0,0,0)$ is Lyapunov stable 
since $q(\bm{x},\bm{z}) \leq 0$ holds. 

We next show that $(\bm{x}(t),\bm{z}(t), \bm{\lambda}(t) ) \to (0,0,0)$, $t \to \infty$.
Since $Q_{\rm all}$ is negative definite, the inequality (\ref{prooflast}) implies that for some positive constant $\theta$
\begin{align}
	&- \theta \int_0^t \|\bm{x}(\tau)\|^2  + \| \bm{z}(\tau)\|^2 {\rm d}\tau \geq \int_0^t q(\bm{x}(\tau),\bm{z}(\tau)) {\rm d}\tau\nonumber\\
	& \geq {\cal V}(\bm{x}(t),\bm{z}(t),\bm{\mu}(t),\bm{\lambda}(t)) - {\cal V}(\bm{x}(0),\bm{z}(0),\bm{\mu}(0),\bm{\lambda}(0))\nonumber\\
	&  \geq - {\cal V}(\bm{x}(0),\bm{z}(0),\bm{\mu}(0),\bm{\lambda}(0)).\label{prooflast2}
\end{align}
Note here that the Lyapunov stability implies the boundedness of
$(\bm{x}(t),\bm{z}(t),\bm{\mu}(t),\bm{\lambda}(t))$.  
Consequently, $(\dot{\bm{x}}(t),\dot{\bm{z}}(t))$, which satisfy (\ref{plant}) and (\ref{pdprop1}), are bounded as well.
From Barbalat Lemma \cite{Khalil_02}, the inequality (\ref{prooflast2}) implies 
 $(\| \bm{x}(t)\|, \|\bm{z}(t)\|) \to (0,0)$, $t \to \infty$.
Finally, recalling that $\alpha$ is positive, we show from (\ref{pdprop3}) that 
$\|\bm{\lambda}(t)\| \to 0$, $t \to \infty$.
}

\end{document}